\documentclass[12pt,oneside,onecolumn,floatfix]{article}
\usepackage{epsfig}
\usepackage{amsmath,amssymb,amsfonts}
\usepackage{a4wide}
\usepackage[top=30pt,bottom=30pt,left=48pt,right=46pt]{geometry}
\usepackage{slashed, enumitem}
\usepackage[utf8]{inputenc}
\usepackage{jheppub}
\usepackage{enumerate}
\usepackage{float}
\usepackage{subfigure}
\usepackage{multirow}
\usepackage{slashed}
\usepackage{placeins}
\usepackage{wasysym} 
\usepackage{mathrsfs} 
\usepackage{caption}
\usepackage{amsmath}
\usepackage{amssymb}
\usepackage{graphicx}
\usepackage{slashed}

\newcommand{\newc}{\newcommand}
\newc{\wt}{\widetilde}
\newc{\ra}{\rightarrow}

\usepackage{array}
\newcolumntype{L}[1]{>{\raggedright\let\newline\\\arraybackslash\hspace{0pt}}m{#1}}
\newcolumntype{C}[1]{>{\centering\let\newline\\\arraybackslash\hspace{0pt}}m{#1}}
\newcolumntype{R}[1]{>{\raggedleft\let\newline\\\arraybackslash\hspace{0pt}}m{#1}}

\def \beq{\begin{equation}}
\def \eeq{\end{equation}}
\def \bea{\begin{eqnarray}}
\def \eea{\end{eqnarray}}
\def \ba{\begin{array}}
\def \ea{\end{array}}
\graphicspath{{./Images/}}

\title{Dark matter capture in celestial objects: Improved treatment of multiple scattering\\ and updated constraints from white dwarfs}

\author{Basudeb Dasgupta,}
\author{Aritra Gupta,}
\author{and Anupam Ray}

\affiliation{Tata Institute of Fundamental Research, Homi Bhabha Road, Mumbai, 400005, India.}

\emailAdd{bdasgupta@theory.tifr.res.in}
\emailAdd{aritragupta@theory.tifr.res.in}
\emailAdd{anupam.ray@theory.tifr.res.in}

\abstract
{We revisit dark matter (DM) capture in celestial objects, including the impact of multiple scattering, and obtain updated constraints on the DM-proton cross section using observations of white dwarfs. 
Considering a general form for the energy loss distribution in each scattering, we derive an exact formula for the capture probability through multiple scatterings.
We estimate the maximum number of scatterings that \emph{can} take place, in contrast to the number \emph{required} to bring a dark matter particle to rest. We employ these results to compute a ``dark" luminosity $L_{\rm DM}$, arising solely from the thermalized annihilation products of the captured dark matter. Demanding that $L_{\rm DM}$ not exceed the luminosity of the  white dwarfs in the M4 globular cluster, we set a bound on the DM-proton cross section: $\sigma_{p} \lesssim 10^{-44} {\rm cm}^2$, almost independent of the dark matter mass between 100 GeV and 1 PeV and mildly weakening beyond. This is a stronger constraint than those obtained by direct detection experiments in both large mass $\left(M \gtrsim 5 \,\,\rm TeV\right)$ and small mass $\left(M \lesssim 10\,\, \rm GeV\right)$ regimes. For dark matter lighter than 350 MeV, which is beyond the sensitivity of present direct detection experiments, this is the strongest available constraint.}

\preprint{TIFR/TH/19-20}


\begin{document}
\maketitle
\vspace{-0.2cm}
\section{Introduction}
\label{sec:intro}
A weakly interacting massive particle (WIMP) is a well-motivated candidate for dark matter --- a scenario that can be tested in a variety of different ways \cite{Bertonebook}. Theories that address the relative smallness of the electroweak scale can ``miraculously'' predict a relic WIMP density that is consistent with the observed cosmological dark matter density~\cite{Steigman:2012nb}. However, in addition to the so-called WIMP miracle, it is the eminently testable nature of WIMPs that has driven the experimental search for said particles. They are generically predicted to have non-negligible interactions with Standard Model~(SM) particles: they can be produced at colliders, can directly collide with SM particles in the lab and elsewhere, and can be indirectly detected through the anomalous fluxes of SM particles from their annihilations.

The very same vaunted testability of WIMPs has however led to some degree of disappointment at not having seen a positive signal yet. Searches using the Large Hadron Collider (LHC) haven't found any trace of new physics up to the TeV scale~\cite{Sirunyan:2017jix,CMS-PAS-HIG-17-023}. As a result, the parent theories now appear to be less well-motivated. The strongest challenge to WIMPs has however come from direct detection experiments that have improved the constraints by many orders of magnitude in the past decade \cite{1805.12562,1608.07648,1708.06917}. For masses around tens of GeV the constraints are now strong enough to disfavor large parts of parameter space motivated by the parent theories. Indirect searches for such dark matter particles have also largely yielded null results \cite{Hooper:2018kfv}. 

Making further progress appears challenging. LHC searches will continue, but not explore significantly higher energies. The more sensitive direct detection experiments will soon reach a scale that will be difficult to improve upon. In addition, they will have to contend with the background due to neutrino-nucleon scattering \cite{1307.5458}, making dark matter searches more difficult. On the indirect detection front, it appears that uncertainties in backgrounds and systematics will continue to plague the attempts to extract a signal for dark matter annihilation.

Nevertheless, it is now being appreciated that the WIMP paradigm is not as constrained as one might naively think. For one, the allowed range of masses for WIMP-like dark matter is larger than previously emphasized. While the hope for new physics at the TeV scale has not yet been met, as far as the WIMP miracle is concerned, the mass range for WIMPs can be quite wide ---  larger than $\sim$ keV, so that the dark matter is cold, but smaller than $\sim100$ TeV, so that its annihilation rate does not violate unitarity. Throughout this mass range, WIMPs can produce the observed cosmological density with a suitable annihilation rate \cite{Feng:2008ya}. Direct detection experiments are not yet sufficiently sensitive at the lower dark matter masses ($\lesssim 1$ GeV) and the possibility of such sub-GeV dark matter remains open \cite{Essig:2011nj,Gunion:2005rw,Dror:2019onn}. Even the upcoming and planned new detectors, will only constrain dark matter heavier than $\sim$ 350 MeV \cite{Petricca:2017zdp}. For even lighter dark matter masses, in the MeV range, electron recoil experiments can be more relevant but their sensitivity is also rather modest \cite{Izaguirre:2013uxa,Essig:2012yx,Abramoff:2019dfb,Abdelhameed:2019szb,Hochberg:2019cyy}. Interestingly, for indirect detection even in the canonical tens-of-GeV range, the perceived stringent constraints are only for annihilations to specific channels and the less model-dependent constraints are not very stringent \cite{Leane:2018kjk}. Obviously, the annihilations to neutrinos are much harder to probe. At larger WIMP masses, the constraints are significantly weaker. Thus, it is worthwhile to re-evaluate the multipronged search strategy for WIMP-like dark matter, recognizing the wider putative range of WIMP masses and unexplored territory.

In this paper, we revisit one prong of this strategy --- the search for signatures of WIMP-like dark matter captured in celestial objects. This search can probe really weak interactions between WIMPs and SM particles, while being practically insensitive to the dark matter mass and annihilation channel. Thus, though the bounds require astrophysical modeling, they are quite strong at low and high masses and are insensitive to many particle physics details. 

\begin{figure}[!t]
	\centering
	\includegraphics[scale=0.45]{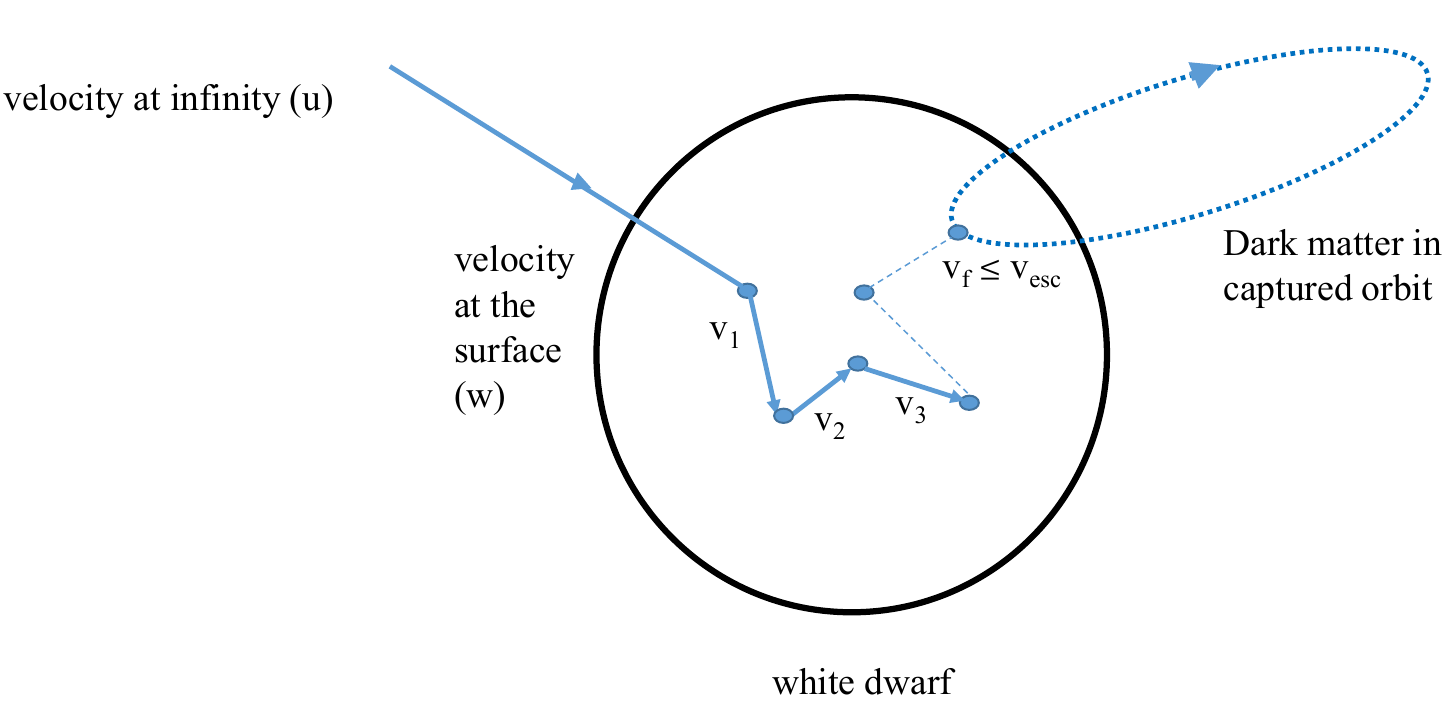}
	\caption{A dark matter coming in from infinity with velocity $u$ enters the celestial object, {\it e.g.}, a white dwarf, with velocity $w$. After this, it scatters one or more times, losing energy, and ultimately its velocity falls below the escape velocity of the white dwarf whence it enters a closed orbit. During its subsequent passages through the star, it will lose more  and more energy before finally being captured.}
	\label{diagram1}
\end{figure}

\newpage
A dark matter particle in the galactic halo, while passing through an astrophysical object, such as the Earth, the Sun, white dwarfs, and neutron stars etc., can lose its kinetic energy by colliding with the protons, neutrons, nuclei, and electrons in the medium. If, as a result, the dark matter particle is slowed to below the object's escape velocity, it gets captured (see Fig.\,\ref{diagram1}). The quantitative description of the capture of dark matter by scattering with nucleons was developed by Press and Spergel \cite {spergel} and by Gould \cite{gould1,gould2,gould3,BBarchealogy}.

These captured dark matter particles have several interesting signatures. Over time, the number density of captured dark matter particles increases within the celestial object, and dark matter may begin to appreciably annihilate. As long as these annihilations are into particles that can thermalize with the medium other details become unimportant and they essentially only heat up the celestial object. Astrophysical observations can be sensitive to such anomalous heating and offer a powerful search strategy.
As an example, neutrino signals in terrestrial neutrino detectors from such captured dark matter within the Sun has been studied in literature earlier ~\cite{Silk:1985ax,Krauss:1985aaa,Freese:1985qw,Hagelin:1986gv,Gaisser:1986ha,Srednicki:1986vj,Griest:1986yu,hep-ph/0009017}. Others have calculated the effect of this accumulated dark matter on cooling of celestial objects \cite{Kouvaris:2007ay,Bertone:2007ae,deLavallaz:2010wp,0705.4298}, or have compared the dark luminosity with the observed luminosity to provide stringent constraints on dark matter interactions with SM particles \cite{Bertone:2007ae, Scott:2008ns,McCullough:2010ai}. More recently, limits on DM-nucleon cross section have also been obtained from non-observation of collapse of massive white dwarfs \cite{Janish:2019nkk} or from neutron star heating \cite{Baryakhtar:2017dbj,Bell:2019pyc,Chen:2018ohx,Camargo:2019wou}.

Most of the earlier treatments assume that the dark matter particle is captured either after a single collision or not at all. This is a reasonable approximation if the cross section of interaction $\sigma$ is small enough, so that the free streaming length $\lambda$ of the dark matter particle is as large as the size of the celestial object itself.
However, this approximation fails in two distinct ways, as recently pointed out by Bramante~{\it et~al.}~\cite{1703.04043}. Firstly, dark matter that is much heavier than the target particles loses small amounts of energy per collision and consequently requires multiple collisions to lose enough energy to be captured. For massive dark matter with mass $\gtrsim 100$ TeV, multiple scatterings can therefore play an important role. Secondly, the smaller the radius of the celestial object the more pronounced will be the effect of multiple scatterings in capturing dark matter. This is understandable because the number of scatterings inside the star is $\sim R/\lambda = n_t\,\sigma\, R \simeq \sigma / R^2$, where $n_t$ is the number density of the target particles inside the object. Obviously larger cross sections lead to higher probability for multi-scatter capture. However, we should keep in mind that the cross section cannot be arbitrarily large. The maximum allowed cross section is given by the geometrical cross section per target particle $\sigma_{\rm sat}=\pi R^2/ {\cal N}_t$, where ${\cal N}_t$ is the number of target particles in the object. In addition, there is yet another way in which the single scattering approximation fails --- if the differential scattering cross section for the dark matter collisions is forward peaked. Here too, energy loss in a single collision is typically small and the cumulative effect of multiple collision may dominate. In this work, we will not dwell too much on this third possibility but the formalism we will develop here is capable of including this possibility as well.

In this work, we improve the treatment of the multi-scatter capture of dark matter in celestial objects and derive constraints using observed white dwarfs. In Sec.\,\ref{sec:formalism}, we recapitulate the original treatments by Gould \cite{gould1,gould2,gould3,BBarchealogy} and the more recent treatment of capture via multiple scattering by Bramante~et~al.~\cite{1703.04043}. We make conceptual and technical improvements in the underlying formalism, treating the energy loss distribution more precisely. We calculate the rate of capture of dark matter through multiple scatterings and its contribution to the luminosities of the stars. In Sec.\,\ref{sec:results}, we then follow the treatment of Bertone and Fairbairn \cite{Bertone:2007ae}, and compare the dark luminosity with the luminosity of white dwarfs observed in the M4 globular cluster. With the inclusion of multiple scattering, we find that for very heavy dark matter with masses\,$\gtrsim 5$\,TeV, where multiple scattering is important, we are able to place stronger constraints than were previously obtained. We are also able to place completely new constraints on dark matter lighter than $\sim$ 350 MeV, and improve the present limit on $\sigma_{p}$ for sub-GeV dark matter from direct detection experiments by several orders of magnitude. We finally conclude in Sec.\,\ref{sec:conclu}.

\section{Analytical calculation of capture rate}
\label{sec:formalism}

\subsection{Review of previous treatments}
A dark matter particle in the halo can be gravitationally attracted towards an astronomical object, undergo one or more collisions inside the object, and eventually get captured. A schematic diagram of such a scenario is shown in Fig.\,\ref{diagram1}. Far away from the object, the dark matter particle has a velocity $u$ and when it reaches the surface of the object its velocity increases to $w$, given by
\begin{equation}
w^2 = u^2 + v_{\rm esc}^2\,.
\end{equation} 
The dark matter particle may undergo one or many scatterings as it transits through the object. The velocity of the incoming dark matter particle decreases as a result of these collisions with the target nucleons or electrons in the medium. If eventually its velocity $v_f$ becomes less than the escape velocity $v_{\rm esc}$, it is captured. Here, we are assuming that the constituent particles of the astronomical object are at rest in the frame of the object. That is, they have no thermal motions and the dark matter particle can only lose energy. This is a good approximation when $\frac{1}{2}M_{\rm DM} v_{\rm esc}^2 \gtrsim k_B T$, {\it i.e.}, the dark matter is not too light and the star is not too hot. For example, in a solar mass white dwarf with temperatures of around $10^6$ K, this lower limit is approximately~$M_{\rm DM}\gtrsim 6$~MeV.

The rate of dark matter particle getting captured in the object depends not only on the size of the object and the flux of dark matter particles, but also on the probability of collisions and the probability of incurring energy loss. Therefore, the capture rate takes the form
\begin{eqnarray}
\label{Eq.CN}
C_{\rm tot} = \sum_{\rm N} C_{\rm N} &=& \sum_{\rm N} \underbrace{\pi R^2}_\textrm{area of the object}\times \, \underbrace{p_{\rm N}(\tau)}_\textrm{probability for $N$ collisions} \nonumber\\
&& \times \,\underbrace{n_{\rm DM} \int \dfrac{f(u)du}{u}\,(u^2+v_{\rm esc}^2)}_\textrm{DM flux}\,\,\,\times\underbrace{g_{\rm N}(u)}_\textrm{probability that $v_f \leq v_{\rm esc}$ after $N$ collisions}\,.
\end{eqnarray}
The capture can occur after the $N^{\rm th}$ collision, and the total rate is simply the sum of the rates corresponding to each $N$. Here, $\pi R^2$ is the area of the astrophysical object within which the dark matter particle is captured. $p_N(\tau)$ is the probability of a dark matter particle with optical depth $\tau$ to undergo $N$ scatterings. If we take into account all the incidence angles encoded in the variable $y$, we have 
\begin{eqnarray}
p_{\rm N}(\tau)=2\int_{0}^{1}dy\,\dfrac{ye^{-y\tau}(y\tau)^N}{N!}\,,
\end{eqnarray}
where the optical depth $\tau = 3\sigma {\cal N}_t/(2\pi R^2)$, ${\cal N}_t$ being the total number of targets in the object and $\sigma$ is the DM-target interaction cross section.

The flux of the captured dark matter particles is given by the product of the dark matter number density in the halo, $n_{\rm DM}=\rho_{\rm DM}/M_{\rm DM}$, and their average velocity. The dark matter energy density near the celestial object is denoted by $\rho_{\rm DM}$ and in the Solar vicinity it is taken to be $\sim$ 0.3 GeV\,cm$^{-3}$. However, in other overdense regions of the Universe it can be much higher. $f(u)$ is the velocity distribution function of the dark matter particle, that is usually taken to be a Maxwell Boltzmann (MB) distribution 
\begin{equation}
f_{\rm MB}(v)= \left(\frac{3}{2 \pi \bar{v}^2} \right)^\frac{3}{2} 4\pi v^2 \exp \left[-\frac{3v^2}{2  \bar{v}^2} \right ]\,,
\end{equation} 
with $\bar{v}\sim 287.8$ km s$^{-1}$ being the rms velocity of the distribution. To account for the motion of the Sun with respect to the rest frame of the galaxy, the distribution function in the Sun's rest frame is boosted, and modeled as
\begin{eqnarray}
f_{\rm Sun}(v)= f_{\rm MB}(v)\,e^{-\eta^2}\dfrac{\sinh(2x\eta)}{2x\eta}\,,
\end{eqnarray}
where $x^2 = 3v^2/(2\bar{v}^2)$ and $\eta^2 = 3\tilde{v}^2/(2\bar{v}^2)$, $\tilde{v}\sim 247$ km s$^{-1}$ being the velocity of the Sun with respect to the dark matter halo. To derive analytic results, we use the usual Maxwell-Boltzmann distribution in the next section. However, all the final results have been computed (numerically) using the boosted distribution, wherever applicable. 

The capture probability $g_{\rm N}$, {\it i.e.}, the probability that the final velocity of dark matter after $N$ scatterings becomes less than $v_{\rm esc}$, {\it i.e.}, $v_f \leq v_{\rm esc}$, is given by
\begin{eqnarray}
\label{Eq.gn}
g_{\rm N}(u)&=& \int_{0}^{1} dz_1 \int_{0}^{1} dz_2 ...\int_{0}^{1} dz_{\rm N}\, s_1(z_1)\times s_2(z_1,z_2)...s_{\rm N}(z_1,z_2 ... z_N) \nonumber \\
&\times& \Theta \bigg(v_{\rm esc}- \left(u^2+v_{\rm esc}^2\right)^{1/2} \prod_{i=1}^{N} (1-z_i \beta)^{1/2}\bigg) \,.
\end{eqnarray} 
Here, $z_i$ is a random variate which takes values between 0 and 1 and encodes the energy lost by the dark matter particle in the $i^{\rm th}$ scattering. The kinetic energy that can be lost in a scattering is given by $\Delta E_{\rm max}=z_i\beta E$, where $\beta = (4 M_{\rm DM} M_t)/{(M_{\rm DM}+M_t)^2}$ is the maximum fraction, with $M_t$ being the mass of the target particles. This variable $z_i$ is in fact closely related to the scattering angle in the center of mass frame, {\it i.e.}, $z =\sin^2(\theta_{\rm CM}/2)$, as explained in Appendix\,\ref{Sec:AppA}. Naturally, $g_{\rm N}$ depends on the probability distributions for the scattering angle encoded in $s_i(z_1,z_2,...z_i)$. Here we confine our discussion to the regime where the differential cross section is independent of the scattering angle and hence all $s_i(z_1,z_2,...z_i)=1$. More general choices of $s_i$ can be considered without much more difficulty.

\subsection{Exact formula for capture probability}
In order to get captured, the final velocity of dark matter particle must become less than the escape velocity. The probability that the dark matter particle with velocity $w$ scatters to a final velocity $v_f$ which is less than or equal to $v_{\rm esc}$, after $N$ number of scatterings, is given by

\begin{equation}
\label{gn}
g_{\rm N}(u)= \int_{0}^{1} dz_1 \int_{0}^{1} dz_2 ...\int_{0}^{1} dz_{\rm N}\, \Theta \bigg(v_{\rm esc}- \left(u^2+v_{\rm esc}^2\right)^{1/2} \prod_{i=1}^{N} (1-z_i \beta)^{1/2}\bigg)\, ,
\end{equation} 
where the $dz_i$ integrals correspond to sum over all possible scattering trajectories. 
We compute this integral analytically to find 
\begin{equation}
g_{\rm N}(u)= \frac{1}{\beta} \frac{v_{\rm esc}^2}{u^2+v_{\rm esc}^2} \left[\frac{1}{\beta} \log \frac{1}{1-\beta} \right]^{N-1}-\left(  \frac{1}{\beta}-1\right)\,.
\end{equation}
 
We interpret $g_{\rm N}(u)$ as the probability that a dark matter particle with speed $u$ at infinity will get captured at its $N^{\rm th}$ collision provided that $N$ collisions occur. See Appendix\,\ref{Sec:AppA} for a brief motivation behind this interpretation. To ensure that $g_{\rm N}$ is positive, we write it as
 \begin{eqnarray}\label{aab}
 g_{\rm N}(u)&=& \left[ \frac{1}{\beta} \frac{v_{\rm esc}^2}{u^2+v_{\rm esc}^2} \left[\frac{1}{\beta} \log \frac{1}{1-\beta} \right]^{N-1}-\left(  \frac{1}{\beta}-1\right) \right ] \times \nonumber \\
 && \Theta \left(  \left[ \frac{1}{\beta} \frac{v_{\rm esc}^2}{u^2+v_{\rm esc}^2} \left[\frac{1}{\beta} \log \frac{1}{1-\beta} \right]^{N-1}-\left( \frac{1}{\beta}-1\right) \right ] \right)\,.
 \end{eqnarray}
This differs from the analogous expression in the previous work, where $z_i$ was replaced by its average value of 1/2 \cite{1703.04043}, which instead gave 
 \begin{equation}
 g_{\rm N}^{\rm approx}(u)=\Theta \left(v_{\rm esc}\prod_{i=1}^{N} \left(1-\frac{1}{2}\beta\right)^{-1/2}-(u^2+v_{\rm esc}^2)^{1/2} \right) \,.
 \end{equation}
The $\Theta$ function in Eq.(\ref{aab}) sets an upper limit to the halo velocity $u$ given by
\begin{equation}
\label{max}
 u^2_{\rm max} \leq v_{\rm esc}^2 \left [ \frac{1}{1-\beta} \left(\frac{1}{\beta} \log \frac{1}{1-\beta} \right)^{N-1}-1    \right ]\,.
\end{equation}
\begin{figure}[!b]
	\centering
	\hspace*{-1cm}
	\includegraphics[scale=0.265]{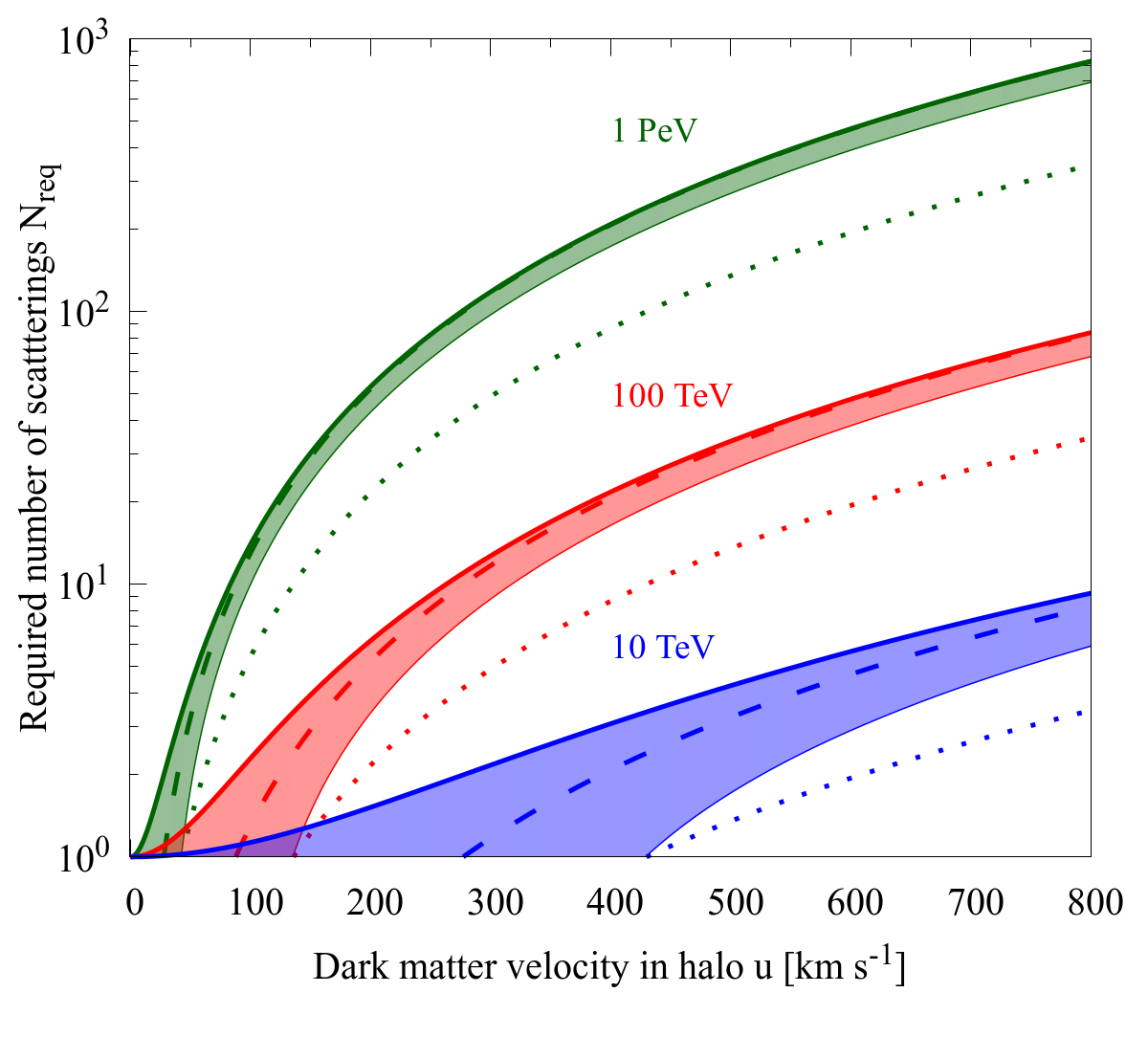}
	\caption{Typical number of scatterings \emph{required} for a dark matter particle with velocity $u$ to be captured by a solar mass white dwarf  with $v_{\rm esc}\sim10^3$\,km s$^{-1}$. The dashed lines are the approximate results where each of the energy loss fractions, $z_i$, was replaced by an ``average'' value of 1/2. The thick solid curves are the maximum number of collisions required for a given $u$ obtained using the exact analytical result. Similarly, the thin lines represent the minimum. The dotted curves represent the absolute minimum number of collisions required for capture, corresponding to the maximum loss in kinetic energy, {\it i.e.}, $z_i=1$.}
	\label{compare}
\end{figure}
This upper limit on $u$ indicates that dark matter particles with arbitrarily large velocity cannot typically be trapped by the celestial object after $N$ scatterings. Furthermore, as $g_{\rm N}(u)$ is a probability, it should also satisfy the condition $g_{\rm N}(u) \leq 1$. This imposes a lower limit on $u$ that was not apparent in the single scattering case where it is trivially satisfied. 
Here, $g_{\rm N}(u) \leq 1$  gives rise to the condition
\begin{equation}\label{min}
u^2_{\rm min} \geq v_{\rm esc}^2 \left [\left(\frac{1}{\beta} \log \frac{1}{1-\beta} \right)^{N-1}-1    \right ]\,.
\end{equation}
This lower limit encodes that if the velocity of the incoming dark matter particle is below this threshold then it is more likely to be captured already {\it before} the $N^{\rm th}$ collision. The expressions for minimum and maximum velocity depend on the assumed expression for $s_i(z_1,z_2,...z_i)$, and the above expressions have been obtained with a uniform distribution, and similar expressions can be obtained for more general choices.
 
\subsection{Number of scatterings required for capture} 
The conditions $0 \leq g_{\rm N}(u)\leq 1$ can also be reinterpreted in a slightly different way. It gives the typical minimum and maximum number of collisions \emph{required} to capture a dark matter particle with a given velocity $u$, 
\begin{equation}
\label{Nreq}
1+ \frac{\log \left[(1-\beta)\frac{u^2+v_{\rm esc}^2}{v_{\rm esc}^2}\right]}{\log \left[ \frac{\log \frac{1}{1-\beta}}{\beta} \right]}\leq  N_{\rm req} \leq 1+ \frac{\log \left[\frac{u^2+v_{\rm esc}^2}{v_{\rm esc}^2}\right]}{\log \left[ \frac{\log \frac{1}{1-\beta}}{\beta} \right]}\,.
\end{equation}
One should not confuse this quantity with the typical maximum number of scatterings that the dark matter \emph{can} experience inside the celestial object before coming to rest. This latter number depends not only on the capture rate in the object but also the life time of the object.

In Fig.\,\ref{compare}, we show the typical maximum required number of scatterings as a function of the dark matter velocity $u$. For smaller dark matter masses and smaller halo velocities, our exact expression in Eq.\,(\ref{Nreq}) (solid lines) is always staying larger than 1 and gives a more meaningful result compared to the approximate result (dashed lines). This is expected, because multi-scatter capture is less viable for light dark matter particles, and the approximation of replacing $z_i$ by its average value of 1/2 is inaccurate for small $N$ \cite{1703.04043}. The improvement for smaller halo velocity $u$ is also understandable on similar grounds. Lower values of $u$ imply a lower initial velocity $w$, and consequently it is more probable for the dark matter particle to get captured after a few scatterings (lower $N$) rather than multiple scatterings. Remarkably, $N_{\rm req}$ is never smaller than 1 according to the result we obtain. The dotted lines in Fig.\,\ref{compare} represent the absolute minimum number of collisions that is essential for the dark matter to be captured from kinematical considerations alone. This happens when the maximum about of kinetic energy is lost in each collision, {\it i.e.}, when $z_i=1$ in Eq.\,(\ref{gn}). Note how the typical minimum number of collisions required (thin lines) is always larger than this absolute minimum number.

\subsection{Capture rate}
\label{Sec:CapRate}

Using the analytical expression for $g_{\rm N}(u)$ in Eq.\,(\ref{aab}), we can now evaluate the capture rate for $N$-scattering. Using energy per unit mass $\zeta = u^2/2$ along with the definition of capture rate in Eq.\,(\ref{Eq.CN}), we find  
 \begin{equation}\label{opp}
   C_{\rm N} = \pi R^2\, p_{\rm N}(\tau) \,n_{\rm DM} \int_{\zeta_{\rm min}}^{\zeta_{\rm max}}  \frac{f(\zeta)d\zeta}{\zeta} (\zeta+\zeta_{\rm esc}) \,g_{\rm N}(\zeta) \,,
 \end{equation}
where $ \zeta_{\rm max}$ and $\zeta_{\rm min}$ can be obtained from Eq.(\ref{max}) and Eq.(\ref{min}) respectively and is given by 
\begin{equation}
  \zeta_{\rm max} = \zeta_{\rm esc} \left [ \frac{1}{1-\beta} \left(\frac{1}{\beta} \log \frac{1}{1-\beta} \right)^{N-1}-1    \right ]\,,
\end{equation}
and
\begin{equation}
   \zeta_{\rm min} = \zeta_{\rm esc} \left [\left(\frac{1}{\beta} \log \frac{1}{1-\beta} \right)^{N-1}-1    \right ]\,.
\end{equation}
with $\zeta_{\rm esc} = v_{\rm esc}^2/2$. 

Finally, using Maxwell Boltzmann distribution from Eq.(\ref{opp}) capture rate for $N$-scattering is 
 \begin{equation}
 C_{\rm N}= \left (\frac{8}{\pi} \right )^\frac{1}{2} \pi R^2\,p_{\rm N}(\tau) \, \frac{n_{\rm DM}}{\sqrt{\bar{\zeta}}} \left [ \frac{\zeta_{\rm esc}}{\beta^N} \left ( \log \frac{1}{1-\beta} \right)^{N-1} p-\left( \frac{1}{\beta}-1 \right) q        \right]\,,
 \end{equation}
where $p$ and $q$ are given as
\begin{equation}
  p= \exp\left[\frac{-\zeta_{\rm min}}{\bar{\zeta}}  \right] -  \exp\left[ \frac{-\zeta_{\rm max}}{\bar{\zeta}} \right]\,,   
\end{equation}
and
\begin{equation}
  q= \left(\bar{\zeta}+(\zeta_{\rm esc}+\zeta_{\rm min}) \right) \exp\left[ \frac{-\zeta_{\rm min}}{\bar{\zeta}}  \right] - \left(\bar{\zeta}+(\zeta_{\rm esc}+\zeta_{\rm max}) \right)  \exp\left[\frac{-\zeta_{\rm max}}{\bar{\zeta}}  \right]\, ,  
\end{equation}
with $\bar{\zeta} = \bar{v}^2/3$.

%%%%%%%%%%%%%%%%%%%
\begin{figure}[!t]
\centering
	\includegraphics[scale=0.27]{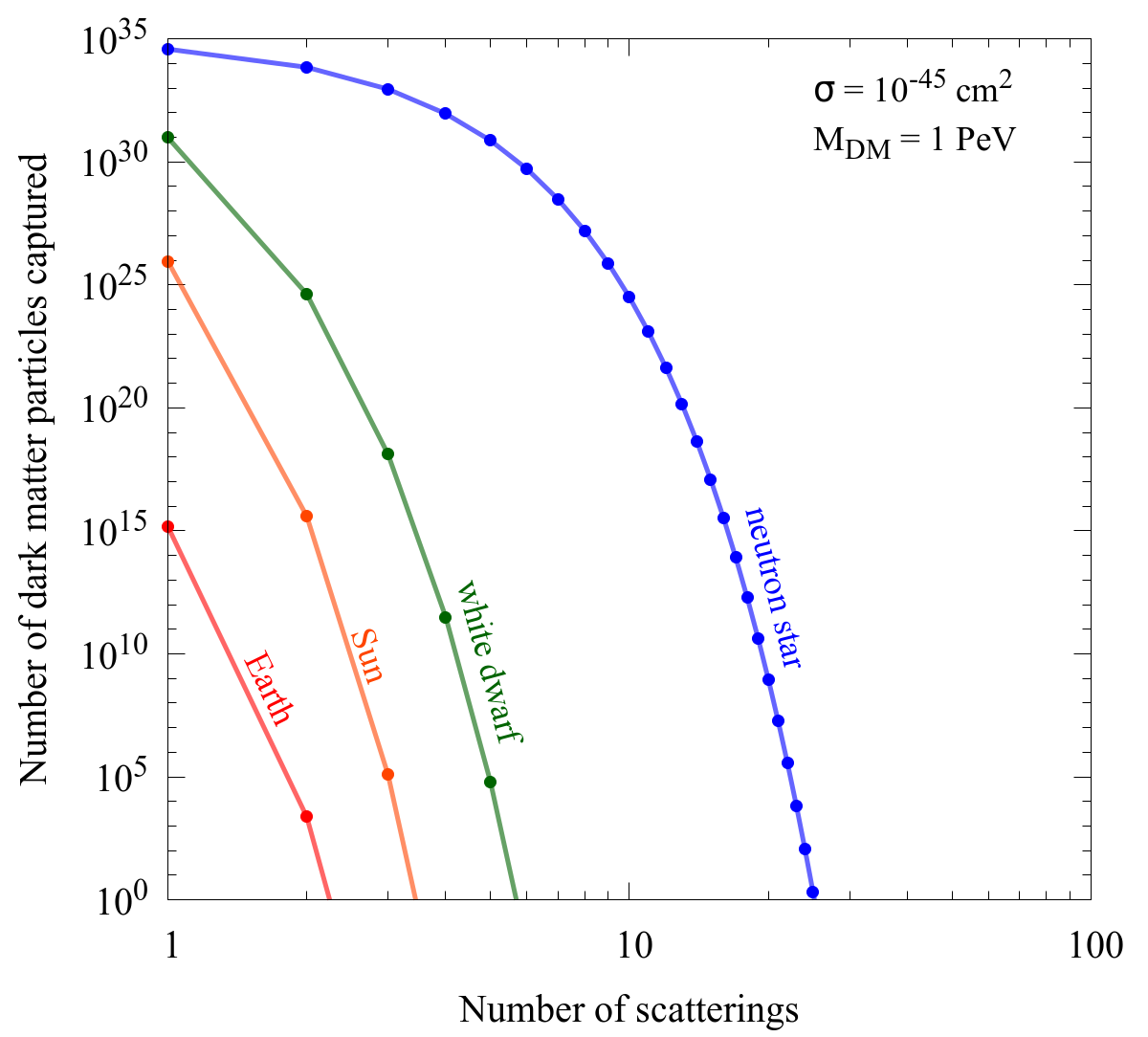}
	\caption{Number of particles captured after $N$ collisions during the life time of the different celestial objects plotted against the number of collisions $N$. Note that, here $\sigma$ denotes the interaction cross section with the relevant target. For example, in Earth the target nucleus is taken to be that of iron while for neutron stars it is simply a neutron. The density of dark matter, for simplicity, has been taken to be the that around the solar system, {\it i.e.}, 0.3 GeV\,cm$^{-3}$.}
         \label{nmaxcorr}
\end{figure}
%%%%%%%%%%%%%%%%%%%

\subsection{Number of scatterings allowed in the object}
It is obvious that the maximum number of scatterings that a dark matter particle can actually undergo must also depend on the time over which such captures can take place.  Roughly, $\tau C_{\rm N}$ gives the total number of dark matter particles that are captured at their $N^{\rm th}$ collision within the lifetime $\tau$ of the celestial object under study. 

In Fig.\,\ref{nmaxcorr}, we show the total number of dark matter particles captured in the Sun, Earth, a typical neutron star and white dwarf, with respect to the number of scatterings it took to capture them within a time $\tau$ taken to be the age of the Universe. Note that the number of captured particles after $N\gtrsim 10$ or so is already smaller than 1, for a cross section $\sigma$ that we will see is marginally allowed. In contrast, the typical maximum number of scatters needed to capture WIMPs, as shown in Fig.\,\ref{compare}, are much larger. This means that the capture rate is dominated by the low-velocity part of the galactic dark matter halo or they are extremely rare events.

It is easy to see that the $C_{\rm N}$ are monotonically decreasing, so that if $\tau\,C_{\rm N}<1$, on average less than one dark matter particles is captured after more than $N$ collisions. Thus, high-$N$ captures are exceedingly rare because the $C_{\rm N}$ are exponentially decreasing with $N$. We use this physically derived criterion to truncate the series in $C_{\rm N}$ where $\tau\,C_{\rm N}=1$.

\section{Results}
\label{sec:results}
\subsection{Luminosity via multi-scatter capture and constraints from white dwarfs}
We now consider the capture of dark matter inside white dwarfs.  White dwarfs are dominantly made up of carbon nuclei, which we take to be the target particle. For the range of dark matter masses that are of interest to us in this work, the typical average momentum transferred to a carbon nuclei inside a solar mass white dwarf is $\lesssim$ MeV. This turns out to be much larger than the inverse of the de Broglie wavelength of the nucleus, which is less than a fm$^{-1}$. Thus, we can treat the relevant collisions to be coherent and elastic. More precisely, the form factor which describes the loss of coherence in case of large energy transfers turns out to be $\sim 1$ for low momentum transfers. For example, using the Helm form factor \cite{PhysRev.104.1466}, we find that for a 10 GeV dark matter $F^2_{\rm Helm} \sim 0.8$ for the maximum possible momentum transfers. For higher dark matter masses, it goes down to $\sim 0.3$ and saturates to this constant value.

To compare with the present direct detection limits, we will translate the DM-carbon cross section $\sigma$ to DM-proton cross section $\sigma_p$. As we are in the regime of coherent scattering, for spin-independent interactions  and assuming equal contributions from protons and neutrons, this translation is simply given by \cite{Lewin:1995rx}

\begin{equation*}
\sigma = \dfrac{\mu_N^2}{\mu_p^2}A^2\sigma_{p}\,.
\end{equation*}
Here, $\mu_N$ and $\mu_p$ are the reduced masses of the dark matter-nuclei and dark matter-proton system. The ratio $\mu_{\rm N}/\mu_{\rm p}$ is $\sim 1$ for light dark matter particles $M_{\rm DM} \lesssim M_p$ and rises to $\sim 12$ for the heavier $M_{\rm DM} \gg M_{\rm carbon}$.
 
The number of captured dark matter particles $N_{\rm cap}$ evolves as $dN_{\rm cap}/dt = C_{\rm tot} - A N_{\rm cap}^2/2$, where $A$ is the annihilation rate of the self-conjugate WIMP. As long as the capture and annihilation processes are in equilibrium\footnote{To ensure that the equilibration time $\leq t_{age} $, the  $\langle \sigma_{a} v \rangle$ must be larger than $\sim 10^{-56}\,\rm cm^{3} \,s^{-1}$ which is obviously much smaller than expected for thermal WIMPs.}, the dark luminosity $L_{\rm DM}$ arising solely from annihilation of captured dark matter particles is given by the mass capture rate $M_{\rm DM}\,C_{\rm tot}$. This additional luminosity is expected to thermalize inside a white dwarf, as long as the annihilation products are SM particles\footnote{For the range of energies considered here, all SM particles, including neutrinos, are expected to thermalize inside a white dwarf.}.
 
\begin{figure}
	\centering
	\includegraphics[scale=0.28]{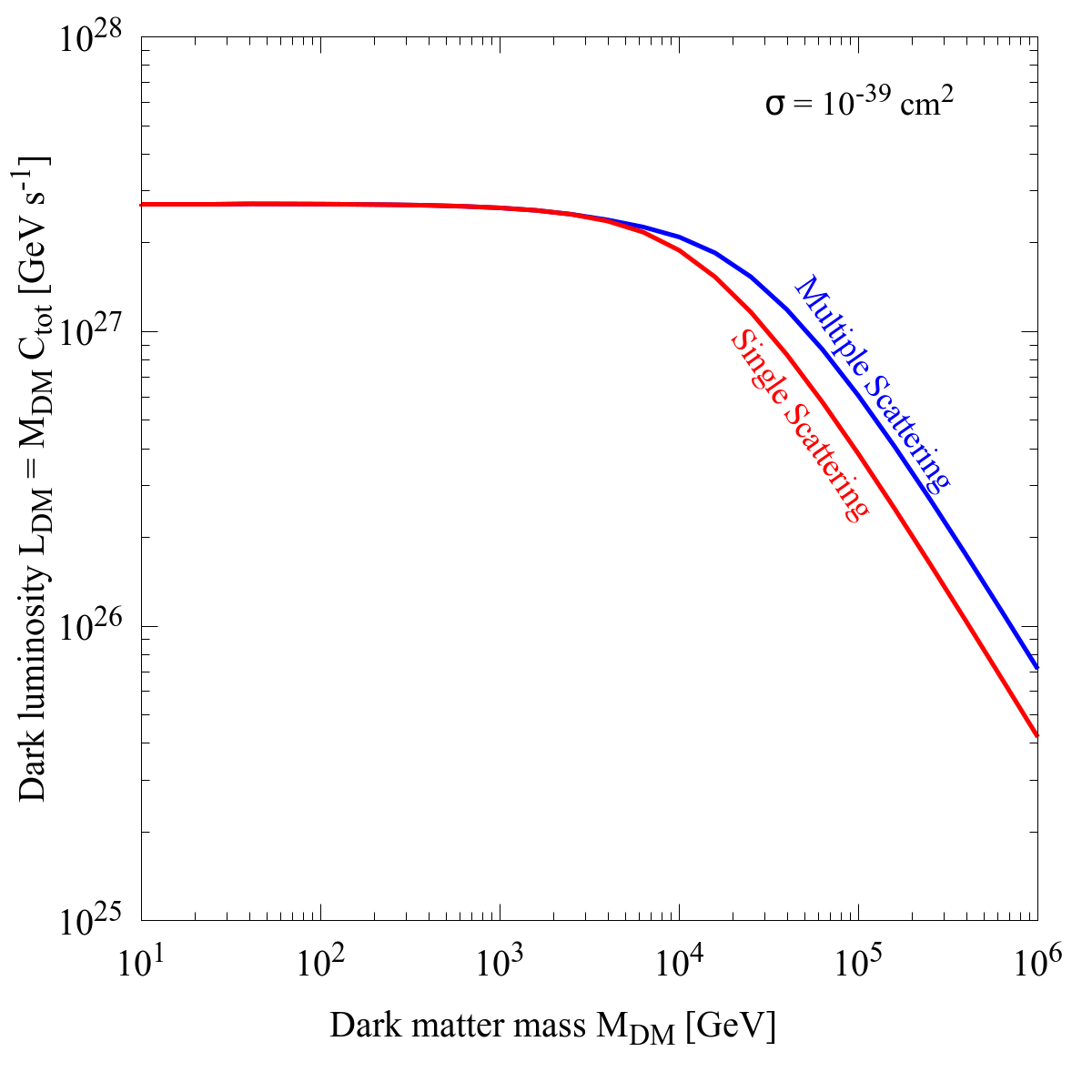}
	\caption{Dark luminosity from annihilation of captured dark matter particles for multiple and single scatterings with carbon nuclei. $\sigma$ denotes the interaction cross section of dark matter with the target.}
	\label{CtotWD}
\end{figure}

In Fig.\,\ref{CtotWD}, we plot the dark luminosity $L_{\rm DM}$ as a function of the dark matter mass. For collision with carbon nuclei inside solar mass white dwarfs, we note that multi-scatter capture becomes important for dark matter masses $\gtrsim 10$ TeV. This is still an order of magnitude below the unitarity bound $\sim 100$ TeV \cite{PhysRevLett.64.615}, and relevant also to canonical thermal WIMPs that are elementary particles.

To obtain constraints on the dark matter interaction cross section, we now compare the dark luminosity $L_{\rm DM}$, which depends mainly on the dark matter properties and the radius of the white dwarf, to the observed luminosities of M4 white dwarfs. McCullough and Fairbairn~\cite{McCullough:2010ai} reported independent measurements of luminosity $L_{\rm obs}$ and temperature  $T_{\rm obs}$ of a few dozen white dwarfs in the M4 cluster. These white dwarfs are unique in that they are among the oldest known celestial objects and are used extensively to study the age of the Universe itself \cite{Richer:1997jk}. In the absence of a dominant burning mechanism inside these dead stars, they are assumed to be nearly perfect black body emitters. Under this assumption, if the luminosity and temperature of a white dwarf are independently measured, we can infer its radius to be $R = \left(L/(4\pi\sigma_0 T^4)\right)^{1/2}$. We next calculate the mass capture rate, {\it i.e.}, $L_{\rm DM}$, using the procedure described in Sec.\,\ref{sec:formalism}, for a fixed dark matter mass and interaction cross section, as a function of the white dwarf radius. Demanding that this dark luminosity should not exceed $L_{\rm obs}$ for a white dwarf of known radius, we impose an upper bound on the dark matter cross section for a given dark matter mass. 

\begin{figure}[!t]
	\centering
	\includegraphics[scale=0.27]{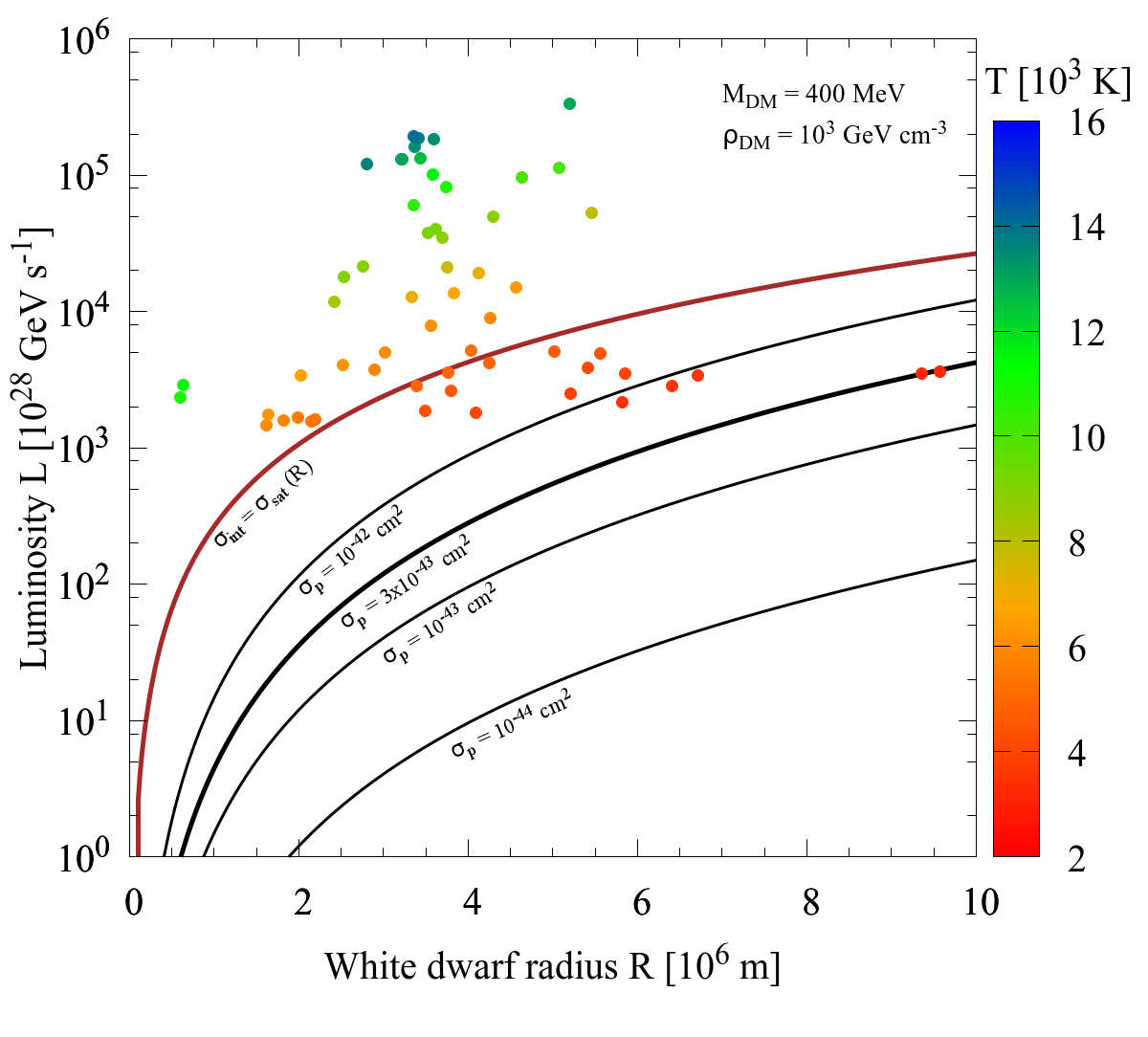}
	\caption{Dark luminosity arising from annihilation of captured dark matter compared with the observed white dwarf luminosities. The dark matter mass was fixed at 400 MeV and five benchmark dark matter-proton cross sections are shown. The topmost curve corresponds to the luminosity when DM-nuclei cross section takes its effectively maximum value, {\it i.e}, $\sigma_{\rm sat}$. The lower curves correspond to smaller cross sections, with the curve marked by $\sigma_p=3\times10^{-43}\,$cm$^2$ being just excluded. The local dark matter density in the M4 cluster is taken to be \mbox{$\sim 10^3$\,GeV cm$^{-3}$ \cite{McCullough:2010ai}} and the dispersion velocity to be $\sim 20$ km s$^{-1}$ \cite{McCullough:2010ai}.}
	\label{result1}
\end{figure}

In Fig.\,\ref{result1}, the solid lines denote the predicted dark luminosity $L_{\rm DM}$ as a function of the white dwarf radius, for several benchmark DM-proton cross sections and a fixed dark matter mass (400 MeV). The position of each colored dot denotes the observed luminosities of a white dwarf and its radius inferred through an independent measurement of its temperature, as explained before. The observed temperature is encoded in color, as per the shown color-bar. The topmost solid line, marked by $\sigma_{\rm sat}$ denotes the maximum attainable \emph{dark} luminosity when the cross section reaches its saturation limit. As argued, $L_{\rm DM}$ must be smaller than the $L_{\rm obs}$. Hence, we find that a DM-proton cross section $\sigma_p \sim 10^{-42}$ cm$^2$ is in tension with the lower luminosity white dwarfs. 

\begin{figure}[!t]
	\centering
	\includegraphics[scale=0.27]{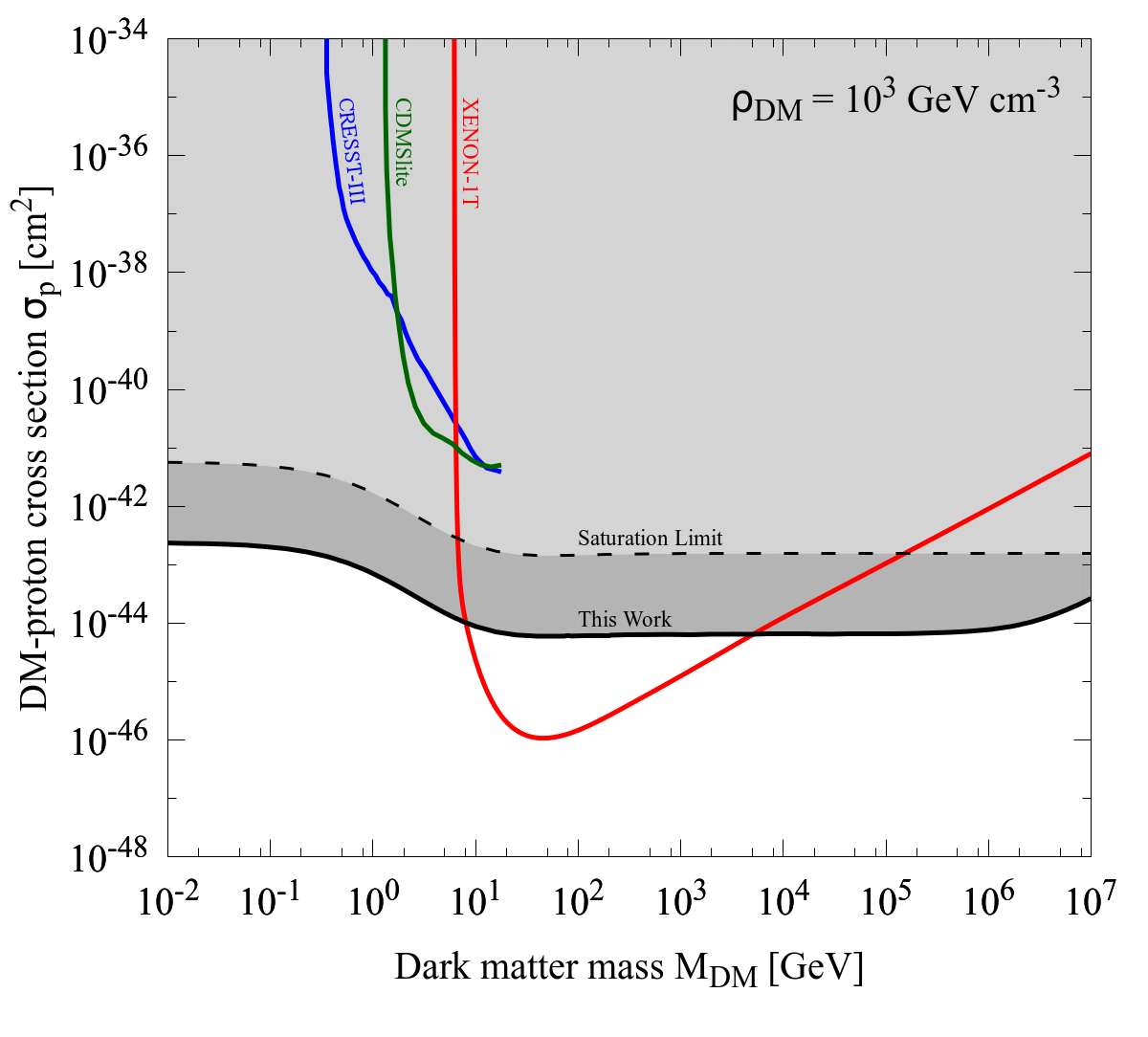}
	\caption{Upper bound on the DM-proton cross section (solid black line) from the observed luminosity of $2.5\times10^{31}$ GeV s$^{-1}$ and a derived radius of $\sim 9\times10^6$ m from a white dwarf in the M4 cluster. Related exclusion limits from direct detection experiments, \texttt{CRESST-III} \cite{Petricca:2017zdp} and \texttt{CDMSlite} \cite{Agnese:2017jvy} in the low mass regime and \texttt{XENON-1T} \cite{Aprile:2018dbl} in the high mass regime, that provide the most stringent bounds. The dashed black line corresponds to $\sigma_{\rm sat}$ (translated to nucleonic cross sections). Above this, in the light gray shaded region, any cross section is essentially equivalent to $\sigma_{\rm sat}$ and ruled out alike.}
	\label{result2}
\end{figure}

In Fig.\,\ref{result2} we furnish an upper bound on $\sigma_p$ as a function of the dark matter mass. This is obtained by demanding that the dark luminosity contribution to the low luminosity white dwarf represented by the right-most red point in Fig.\,\ref{result1} be smaller than its observed luminosity. The observed luminosity of this white dwarf is $\sim$\,$2.5\times 10^{31}$ GeV\,s$^{-1}$. We assume that, in the worst case scenario, all of this luminosity comes only from burning of trapped dark matter inside the star. The radius of this white dwarf is inferred to be $\sim 9\times10^6$ m. The most stringent bounds obtained from different direct detection experiments in the light and heavy dark matter regimes are shown for comparison. Notice that the constraint is practically independent of the dark matter mass and, unlike the corresponding constraint from direct detection experiments, it remains quite strong at lower and higher dark matter masses. This is simply because $L_{\rm DM}=M_{\rm DM}C_{\rm tot}$, while $C_{\rm tot}$ itself scales as $1/M_{\rm DM}$ due to its dependence on dark matter number density. As a result the dark matter mass-dependence cancels out and the constraint is practically mass-independent in this range. The weak mass-dependence of the constraint on the DM-proton cross section $\sigma_p$ is due the presence of the form factor and the ratio of the reduced masses, both of which depend on the dark matter mass. The dark-gray shaded region in Fig.\,\ref{result2} corresponds parameter space excluded by our results. Cross sections exceeding $\sigma_{\rm sat}$ (above the dashed line) are also excluded, but at the same significance as at the dashed line.

The constraints obtained here are highly competitive. In the low-mass regime, {\it i.e.}, below 10 GeV, it is the strongest available bound. For such light dark matter masses, the constraint from direct detection experiments is rather weak and we find that we were able to make an improvement of nearly 3--7 orders of magnitude when compared with \texttt{CRESST-III} \cite{Petricca:2017zdp} or \texttt{CDMSlite} \cite{Agnese:2017jvy}. Crucially, because of the signature mass-independence, one finds stringent bounds for dark matter particles less than 350 MeV that are below the sensitivity of typical direct detection experiments. Likewise, in the high-mass regime above a few TeV, these constraints are the strongest. In this regime the improvement due to multi-scattering is important.

\subsection{Variations on the theme}
It is possible that dark matter particles are leptophilic and thus only collide with electrons, or perhaps have interactions that are not spin-independent. In these scenarios, and several others, the calculation we perform can be repeated to obtain a corresponding constraint, though they are not as strong. As an illustration of how the constraint changes, we rederive our constraint for DM-electron scattering in solar mass white dwarfs. This is also motivated by the fact that multiple scatterings are expected to become more important for much smaller dark matter masses with electrons as targets.

\begin{figure}
	\centering
	\includegraphics[scale=0.27]{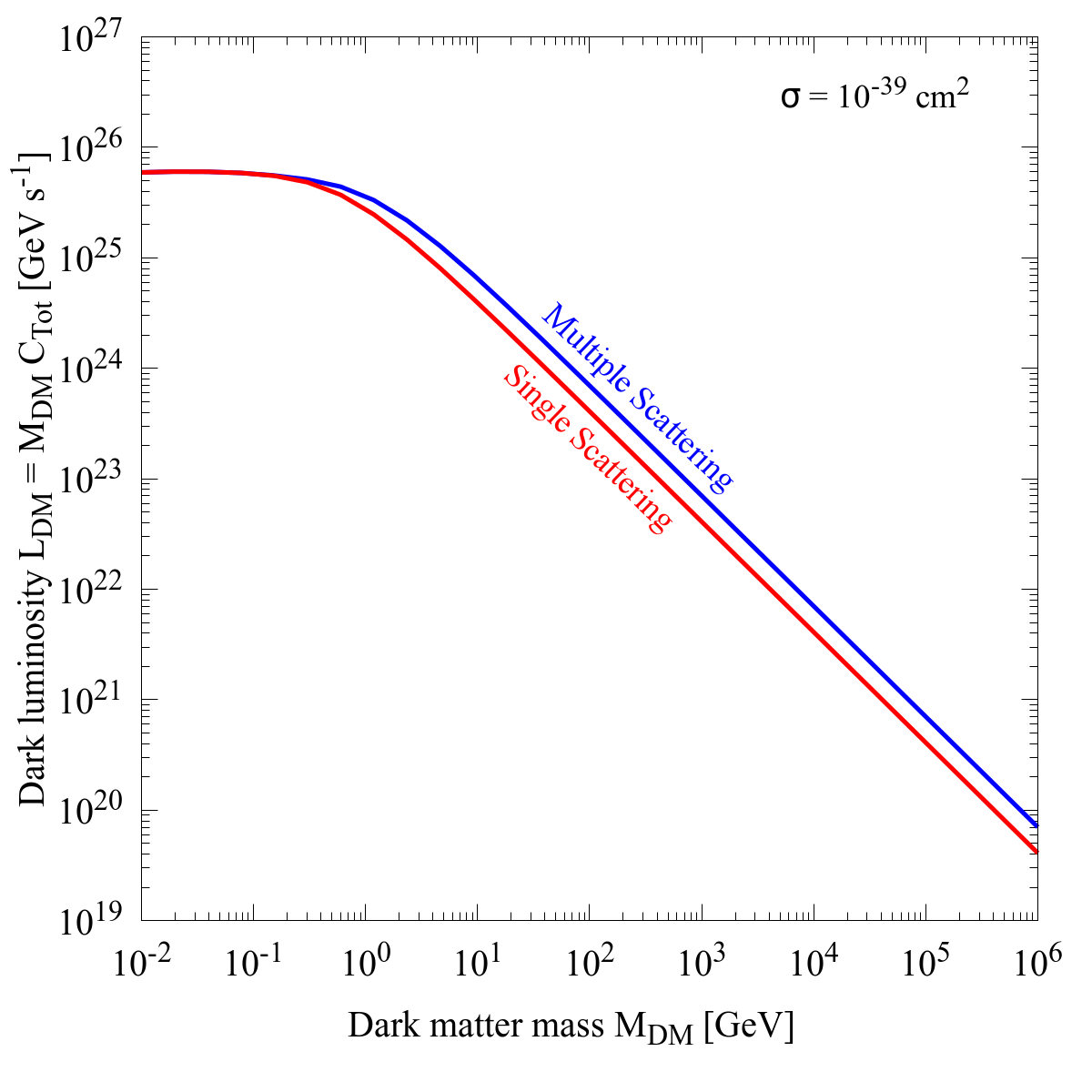}
	\caption{Dark luminosity from annihilation of captured dark matter particles for multiple and single scatterings with electrons. $\sigma$ denotes the total cross section of dark matter with the target electrons.}
	\label{CtotWD2}
\end{figure}

When one considers electrons in a white dwarf, it becomes important to consider the efficiency factor due to Pauli blocking. The electron is pushed to a higher momentum state due to its collision with the incoming dark matter particle. However, this higher state may or may not be available, owing to Pauli exclusion. Hence, while calculating the total capture rate we have to include a corresponding efficiency factor \cite{Goldman:1989nd,McDermott:2011jp}
\begin{eqnarray}
\xi = {\rm Min}\left[1, \dfrac{\delta p}{p_F}\right]\,,
\end{eqnarray}
where $\delta p \sim \sqrt{2}\,\mu_r \,v_{\rm esc}$ with $\mu_r$ being the corresponding reduced mass. The Fermi momentum is $p_F = (3\pi^2\,n_t)^{1/3}$. For a solar mass white dwarf with $R \sim \mathcal{O}(10^{-2})R_{\rm Sun}$, and, for the range of dark matter masses that we consider in this work, we find that for collisions with electrons $\xi \sim 10^{-2}$ but with nucleons it is $\sim 1$. So, we expect a suppression in case of collisions with electrons but not with a nucleon (or other heavier nuclei). The dark luminosity $L_{\rm DM}$ in the case of collision with electrons inside a white dwarf is shown in Fig.\,\ref{CtotWD2}. We see, unlike the case with collisions against nuclei, here multi-scatter capture becomes important for much lighter dark matter masses $\sim \mathcal{O}(1)$ GeV, as expected.
Unfortunately, with electrons as targets, we find that even with the largest allowed cross section, {\it i.e.}, $\sigma_{\rm sat}$, the dark luminosity $L_{\rm DM}$ is always less than the observed luminosity of all the white dwarfs in the M4 globular cluster. Hence, we are not able to constrain any physically relevant cross sections for a large range of dark matter masses. The main source of this suppression in $L_{\rm DM}$ in the case of electrons comes from the efficiency factor due to the Pauli blocking as discussed earlier. If somewhat colder white dwarfs are observed in future they would lead to very strong bounds.

The limits presented in this work concern only with the DM-proton spin-independent cross sections. This is because, the white dwarfs are primarily rich in spin-zero carbon nuclei which are the principal targets for the dark matter particles. To derive similar bounds on DM-proton spin-dependent cross sections, one has to consider capture of dark matter through collisions with targets having a net non-zero spin \cite{Albert:2018jwh}.

Recently several groups have explored scenarios where dark matter is captured inside neutron stars due to its collision with electrons \cite{Bell:2019pyc} and neutrons \cite{Baryakhtar:2017dbj}, and consequently provided stringent projected constraints on $\sigma_{e}$ and $\sigma_{n}$ respectively. The limits we have obtained for the DM-proton cross section are competitive. 

\section{Summary}
\label{sec:conclu}

We have revisited the formalism for capture of dark matter in celestial objects and, upon making improvements to the same, obtained constraints on the dark matter interactions with SM particles. One of the key improvement we have made is a careful consideration of the energy loss in each collision, that we relate to the differential cross section. Further, we have generalized the formalism to be able to include arbitrary energy loss distributions, in contrast to the uniform distribution based on the assumption of a heavy mediator. We then computed the capture probability after $N$ scatterings exactly, that leads to well-behaved results at low dark matter velocities. By studying the analytical results, we were able to interpret the calculation more physically, which provides a clearer picture of the importance of multiple scatterings.

As a concrete improvement, we calculated the dark luminosity of white dwarfs in the M4 globular cluster arising only from the annihilation of captured dark matter. 
With electrons as targets, we found that even with the largest allowed cross section, {\it i.e.}, $\sigma_{\rm sat}$, the dark luminosity is well below the observed luminosities. Thus, in order to obtain a constraint, one would need to either model these stars accurately and estimate the non-dark contribution, or find much colder white dwarfs. The main source of this suppression in $L_{\rm DM}$ in the case of electrons comes from the efficiency factor due to the Pauli blocking. More encouragingly, with carbon nuclei as targets, this suppression is absent. We were thus able to place a constraint on the DM-proton (or equivalently DM-nucleon) cross section $\sigma_p$ that is stronger than direct searches. The improvement occurs mostly in the light (up to 7 orders of magnitude) and heavy dark matter regions ($\sim 1$ order of magnitude). As a bonus, we found that our constraints can be extended to lower dark matter masses ($\lesssim 350$ MeV), where there are no existing bounds from terrestrial direct detection experiments. These bounds at lower masses are much stronger than the recently reported constraints on very light dark matter due to their interactions with cosmic rays \cite{Cappiello:2018hsu,Bringmann:2018cvk}, though with very complementary systematics.  As caveats, we must note that the constraint is strongly dependent on the capture of low-velocity dark matter particles and thus subject to the uncertainties in the velocity distribution of dark matter in the M4 cluster. Microstructure in the dark matter density and velocity, {\it e.g.}, due to possible dark matter streams or disks, might affect these constraints strongly.

\section{Note added}
The formalism presented in this paper was mainly intended to be applied for DM-target cross sections $\sigma$ that are within the optically thin regime. The motivation for inclusion of multiple collisions is not so much that $\sigma$ is large, but rather that the DM particle is massive and loses little energy in any one collision or that the star is compact~(cf.~Sec.\,\ref{sec:intro}).  For $\sigma> \sigma_{\rm sat}$, as evident from Fig.\,\ref{result2}, we used a redefinition of $\sigma = {\rm min}[\sigma,\sigma_{\rm sat}]$ which gives a capture rate that monotonically increases with $\sigma$ and saturates at a value obtained for $\sigma=\sigma_{\rm sat}$. Quantitatively though, even with this redefinition, the asymptotic capture rate (for $\sigma\to\infty$) does not approach the transit rate (which it should); this regime has been treated with greater accuracy in subsequent papers, e.g., in arXiv:2404.16272.

\section*{Acknowledgements}
We thank John Beacom, Sudip Bhattacharyya, Francesco Capozzi, Sudip Chakraborty, Sourav Chatterjee, Anirban Das, Subhajit Ghosh, and Georg Raffelt for many useful suggestions and discussions.
The work of B.D. is partially supported by the Dept. of Science and Technology of the Govt. of India through a Ramanujan Fellowship and by the Max-Planck-Gesellschaft through a Max-Planck-Partnergroup.

\appendix

\section{Kinematics and energy loss in one or more collisions}
\label{Sec:AppA}
The kinematics of single elastic scattering dictate that the fractional energy loss $\Delta E/E$ is restricted in the range 
 \begin{equation}\label{Eq:total}
0 \leq \frac{\Delta E}{E} \leq \beta\,,
 \end{equation} 
where 
  \begin{equation}
 \beta = \dfrac{4 M_{DM} M_t}{(M_{DM}+M_t)^2}
 \end{equation}
is the maximal energy loss fraction that itself is $\leq1$.
 
On the other hand, scattering to velocity $ v_{\rm esc} $ or less requires a minimum energy loss
 \begin{equation}
 \label{Eq:interval}
\frac{\Delta E}{E} \geq \frac{w^2-v_{\rm esc}^2}{w^2} =\frac{u^2}{u^2+v_{\rm esc}^2}\,.
 \end{equation}
 Eq.(\ref{Eq:total}) can be rewritten as
\begin{equation}
\Delta E=\beta E \cos^2\theta_{\rm recoil},
\end{equation}
where the recoil angle $\theta_{\rm recoil}$ is related to the scattering angle in CM frame $\theta_{\rm CM}$ by
\begin{equation}
\theta_{\rm recoil}=\frac{\pi}{2} -\frac{\theta_{\rm CM}}{2}\,.
\end{equation}

We define the collision parameter
$z=\cos^2\theta_{\rm recoil}$ which takes values in the range $[0,1] $. If we denote the velocity after collision by $v_f$, then, from the kinematics described above, we get 
\begin{equation}
v_f= (1-z \,\beta)^{1/2}\,\left(u^2+v_{\rm esc}^2\right)^{1/2}\,.
\end{equation}
A simple extension of this result leads us to the expression of $v_N$, the velocity after $N$ collisions. It is given by $v_N= \prod_{i=1}^{N} (1-z_i \beta)^{1/2}\left(u^2+v_{\rm esc}^2\right)^{1/2}$, with $z_i$ being the collision parameter for the $i^{\rm th}$ scattering.

The distribution of $z$ is determined by the distribution of $\theta_{\rm CM}$, which in turn is dictated by the differential cross section ${d\sigma}/{d\Omega}$ of the relevant scattering process,
\begin{eqnarray}
s(z)=\dfrac{1}{\sigma}\dfrac{d\sigma}{d\Omega}\,,
\end{eqnarray}
where $\Omega$ is the solid angle with $d\Omega = \sin\theta \, d\theta \, d\phi$. 
As an example, consider a fermionic dark matter with mass $M_{\rm DM}$ whose interaction is mediated by a vector or a scalar of mass $M_{\rm med}$. In the non-relativistic perturbative limit, the Born differential cross section of dark matter self interaction is given by
\begin{equation}
\dfrac{d\sigma}{d\Omega_{\rm CM}} = \dfrac{\alpha_D^2 M_{\rm DM}^2}{\left(
	M_{\rm DM}^2 v_{\rm rel}^2 \sin^2(\theta_{\rm CM}/2) + M_{\rm med}^2\right)^2}\,,
\end{equation} 
where $\alpha_D$ is the interaction strength. When the mediator is much heavier than dark matter, the differential cross section is approximately a constant with respect to the scattering angle. In such scenarios, $s(z)$, {\it i.e.}, the distribution of $z$, is uniform. In the opposite limit of a very light mediator, where ${d\sigma}/{d\Omega_{\rm CM}} \sim {1}/{\sin^4(\theta_{\rm CM}/2)}$, the assumption of uniform distribution function is a poor approximation. In this case, the distribution of $\cos^2\theta_{\rm recoil} \equiv z $ goes as $1/z^2$.

If the distribution of energy loss is uniform, as in the case for a massive mediator, then using Eq.(\ref{Eq:total}) and Eq.(\ref{Eq:interval}) the probability for the dark matter particle to scatter to a velocity $v_{\rm esc}$ or less turns out to be 
  \begin{equation} 
  \label{Eq:g1}
 g_1(u)= \frac{1}{\beta} \left(\beta-\frac{u^2}{u^2+v_{\rm esc}^2} \right) \Theta \left(\beta-\frac{u^2}{u^2+v_{\rm esc}^2} \right) \,.
 \end{equation}
The $\Theta$ function ensures the positivity of this probability and sets an upper limit on the halo velocity $u$. This is understandable because a dark matter particle with an arbitrarily large halo velocity cannot lose enough energy to get captured after a single collision. The remainder of the expression has a simple interpretation: it is the range of energy loss that leads to a successful capture, divided by the range of possible energy loss. For a uniform distribution of the energy loss, this ratio is the probability that there is sufficient energy loss that leads to a capture.
   
Eq.(\ref{Eq:g1}) can also be looked upon as a special case of the more general expression of $g_{\rm N}(u)$ presented in Eq.(\ref{Eq.gn}), {\it i.e.},
\begin{eqnarray}
g_1(u)=\int_{0}^{1}\,dz\,\Theta\left(v_{\rm esc}-(1-z\beta)^{1/2}(u^2+v_{\rm esc})^{1/2}\right)\, .
\end{eqnarray}
This, when integrated, yields Eq.(\ref{Eq:g1}) as expected. Furthermore, if we use $N=1$ in the general expression for the capture rate $C_N$ as given in Eq.(\ref{Eq.CN}), and use the fact that $p_1(\tau) \sim \, 2\tau/3$ for $y\,\tau \ll 1$ along with the definition of the optical depth $\tau$, we find that $\pi R^2 p_1(\tau) \rightarrow \sigma {\cal N}_t$, where ${\cal N}_t$ is the total number of targets present within the celestial body. Eq.(\ref{Eq.CN}) thus reduces to 
\begin{equation}
\label{c1}
C_1 =  \sigma \,{\cal N}_t \int \frac{f(u)du}{u}\,(u^2+v_{\rm esc}^2) \,g_1(w)\,
\end{equation}
Therefore, we recover the familiar result for single scatter capture as a limiting case of the general framework of capture through multiple scatterings, as presented here.
\bibliographystyle{JHEP}
\bibliography{ref}
\end{document}